# Ferroic properties of Fe-doped and Cu-doped $K_{0.45}Na_{0.49}Li_{0.06}NbO_3$ ceramics

Laijun Liu[*a], Danping Shi[a], Longlong Fan[b], Jun Chen[b], Guizhong Li[a], Liang Fang[a], Brahim Elouadi[c]

[a]*Key laboratory of Nonferrous Materials and New Processing Technology, Ministry of Education, College of Materials Science and Engineering, Guilin University of Technology, Guilin 541004, China*
[b]*Department of Physical Chemistry, University of Science and Technology Beijing, Beijing 100083, China*
[c]*Laboratory of Chemical Analysis Elaboration and Materials, Engineering (LEACIM), Université de La Rochelle, Avenue Michel Crépeau, 17042 La Rochelle, Cedex 01, France*

## Abstract

Ferroelectric and magnetic properties of Fe-doped $K_{0.45}Na_{0.49}Li_{0.06}NbO_3$ (KNLN) and Cu-doped KNLN ceramics were investigated. Comparing with pure KNLN ceramic, Fe-doped KNLN shows ferromagnetic and ferroelectric properties simultaneously while Cu-doped KNLN shows stronger diamagnetic and ferroelectric properties. The magnetic property in KNLN matrix are believed to be induced by point defects. The ferromagnetism of Fe-doped KNLN can be explained by the F-center exchange mechanism, while the antiferroelectric-like loop and diamagnetism of Cu-doped KNLN was explained by the special defect structure and $Cu^{1+}$ ion at A site.

---

[*] Corresponding author: Tel.: +867735893395; fax: +867735896290. Email: ljliu2@163.com (L. Liu)

1. **INTRODUCTION**

Much attention has been focused on perovskite oxides ($ABO_3$) due to their relatively simple structure and many excellent physical properties [1], such as superconductivity, giant dielectric permittivity, giant magnetoresistance and ferroelectricity. Ferromagnetic materials are used in the fields of information storage, while for ferroelectric materials, the switchable polarization under electric field makes them useful in non-volatile memories. The application of ferroelectric materials could be greatly extended [2] if the ferroelectric performance is coupled with magnetic field, which are called "multiferroics". Materials that are simultaneously ferroelectric and ferromagnetic are gaining more and more attention.

However, very few single-phase bulk materials have been found to be multiferroic at room temperature. Most perovskite ferroelectrics include transition metal ions with empty $d$ shells, such as $Nb^{5+}$ and $Ti^{4+}$ at B site, which seem to be a prerequisite for ferroelectricity [3]. On the contrary, a magnetism requires transition metal ions with partially filled $d$ shells, such as $Mn^{3+}$ and $Fe^{3+}$ [4]. In perovskite oxide $BiFeO_3$ [4,5], the stereochemically active 6s lone-pair of $Bi^{3+}$ ions contribute to ferroelectricity, and interacting spins of the $d$-electrons of $Fe^{3+}$ ions gives rise to the magnetism. In addition, the ferroelectricity associates with the geometry of its structure [6] in manganites, for example, $YMnO_3$.

In recent years, many researchers have focused on the substitution of magnetic ions at B site to induce magnetism in perovskite ferroelectrics, such as $PbTiO_3$ [7], $BaTiO_3$ [8,9] and $K_{0.5}Na_{0.5}Nb_{0.95}Ta_{0.05}O_3$ [10] by created of oxygen vacancies. However, the influence of defect characteristic on the coupling between magnetism (or diamagnetism) and ferroelectricity is rarely considered. This motivate us to study the influence of point defects (substitution ions or oxygen vacancy) on the magnetism and ferroelectricity in perovskite compounds. The selected material is $K_{0.45}Na_{0.49}Li_{0.06}NbO_3$ (KNLN), a lead-free piezoelectric material with excellent piezoelectric and ferroelectric properties [11]. Here, we report the observation of ferromagnetism and diamagnetism at room temperature as well as ferroelectricity in iron doped KNLN and copper doped KNLN ceramics, respectively.

## 2. EXPERIMENTAL PROCEDURE

$K_{0.45}Na_{0.49}Li_{0.06}NbO_3$ (KNLN), $K_{0.45}Na_{0.49}Li_{0.06}NbO_3$ doped 0.5mol% $Fe_2O_3$ (KNLN-Fe) and $K_{0.45}Na_{0.49}Li_{0.06}NbO_3$ doped 1mol% CuO (KNLN-Cu) ceramics were prepared by the conventional solid reaction method. High-purity (99.99%) $K_2CO_3$, $Na_2CO_3$, $Li_2CO_3$, $Nb_2O_5$, $Fe_2O_3$ and CuO were used as raw materials. The weighted materials were milled with Y-doped $ZrO_2$ balls for 8h using ethanol as a medium and then calcined at 850ºC for 2h. The calcined powder was dried and milled again for 8h, and then pressed into pellets of 12mm in diameter. The pellets were sintered at a temperature range from 1020 to 1150 ºC for 2h.

Bulk densities were determined by the Archimedes immersion technique and shown in Table 1. Phase identification of the ceramics was examined by X-ray diffraction (XRD; PANalytical X'pert PRO MPD) with Cu $K_{\alpha 1}$ radiation. Samples were polished and painted with silver paste on both sides, and then fired at 550ºC for 30 min. The room-temperature ferroelectric properties at 1 Hz were measured using the ferroelectric testers (aixACCT: TF-2000). Small bulks ($1 \times 1 \times 1$ mm$^3$) were cut from the as-prepared samples for magnetic property measurements and magnetization-field curves were achieved using physical property measurement system (PPMS; Quantum Design). The capacitor and conductivity with or without applied magnetic field were measured using Precision LCR Meter at room temperature. X-Band (9.4 GHz) continuous-wave and pulsed EPR measurements were performed on a Bruker ElexSys 680 spectrometer, using a cylindrical $TE_{011}$ dielectric ring resonator (Bruker). The exact magnetic field values were calibrated by a standard field marker (polycrystalline DPPH with g = 2.0036). All spectra were recorded at a temperature of 90 K, using a helium-flow cryostat (Oxford).

## 3. RESULTS AND DISCUSSION

XRD patterns of the KNLN, KNLN-Fe and KNLN-Cu ceramics are shown in Fig. 1. Both doped samples possess a single perovskite structure with orthorhombic symmetry and no impurity can be found, indicating that most of iron and copper ions diffuse into the host lattice. Since all samples were sintered in air, Fe ions should maintain their raw valence of 3+[12] while Cu ions should be 2+, which was proved by many researchers using electron paramagnetic resonance (EPR)[13]. The ionic radii of $Fe^{3+}$ (0.55 Å for LS and 0.645 Å for HS) $Cu^{2+}$ (0.73 Å) ions are very similar to the $Nb^{5+}$ ion (0.64 Å) and much smaller than that of the $Na^+$ ion (1.39 Å) and $K^+$ ion (1.64 Å), therefore, it is reasonable to imply that transition metals ions substitute into the Nb site of KNLN lattice. The structure of the ceramics was refined by using the XRD Rietveld method supported by the software TOPAS[14]. Since the perovskite type subcell of $K_{0.5}Na_{0.5}NbO_3$ is monoclinic while its unit cell has orthorhombic symmetry at room temperature. Therefore, the crystal structure of KNLN can be described in the monoclinic symmetry instead of orthorhombic one[15]. The cell parameter and cell volume were obtained (shown in Table 1) according to *Pm* (group 6) space group [16]. KNLN-Fe has the maximum cell volume instead of KNLN-Cu. A possible reason is that $Fe^{3+}$ occupied B site compensates intrinsic oxygen vacancy and restrain the evaporation of K/Na, while higher concentration of oxygen vacancy decreases the cell volume. Although the compounds structure was refined by Rietveld method, the exact amount of Fe and Cu in the KNLN ceramics did not be known due to the limit of our X-ray diffractometer and very small amount addition of Fe and Cu.

The microstructure of thermally etched ceramics was investigated by scanning electron microscope (SEM, JEOL JSM-6380LV). Our studies concentrated on the effect of dopants on the final microstructure. SEM images of KNLN-based ceramics with different dopants are presented in the inset of Fig. 1. An average grain size of 10–15 μm was obtained for these samples by using the linear intercept method (shown in Table 1). Homogeneity varies slightly between the different compositions while KNLN-Fe shows a cuboid morphology. However, the difference between the homogeneity and grain size of the samples is not significant enough to conclude if there is any influence of composition. The fine-grained microstructure is quite uniform with few grain-boundary porosities in KNLN-Cu, which is associated with the defect structure, such as oxygen vacancies. The EDS measurement was employed to detect the element distribution for the samples, as shown in Fig. 2. No obvious difference for the EDS spectrum between them is observed although trace of Cu or Fe can be detected in Fig. 2b and 2c, respectively. Lithium is not observed in the EDS spectrum because it has too low of an atomic number to be detected with EDS.

Polarization hysteresis loops of the KNLN, KNLN-Fe and KNLN-Cu ceramics were measured at room temperature and 1 Hz, as shown in Fig. 3. Their curves show saturation polarization when applied electric field is 4 kV/mm. The electric coercive field ($E_c$) and remanent polarization ($P_r$) of the KNLN ceramics are 0.95 kV/mm and 20.6 μC/cm$^2$ respectively. After the addition of 1 mol% iron element, the loop becomes slim with $E_c$= 1.20 kV/mm, $P_r$ =10.7 μC/cm$^2$. Unlike KNLN and KNLN-Fe, the KNLN-Cu ceramic exhibits a double $P$-$E$ loop at 1 Hz, which associates with the symmetry-conforming principle of point defects [17]. The increase in $E_c$ and decrease in $P_r$ indicate that $Fe^{3+}$ and $Cu^{2+}$ act as an acceptor in KNLN ceramics, which is consistent with ref. 12 and 13. Because the valency of iron and copper is lower than that of niobium at B site, oxygen vacancies shared with nearby unit cells will be formed in order to maintain the charge neutrality. The oxygen vacancies pin the domain wall motion thus weaken the ferroelectric properties.

Magnetic properties of the KNLN-Fe and KNLN-Cu ceramics were measured at different temperatures. The KNLN-Fe ceramic shows ferromagnetism with hysteresis loops shown in Fig. 4(a1), confirming the existence of ferromagnetic long-range order. The temperature dependence of magnetization for the KNLN-Fe ceramic at zero-field cooled (ZFC) and 0.5T field cooled (FC) conditions in the temperature range from 15 to 390K is shown in Fig. 4(a2). A sharp decrease of magnetization with increasing temperature can be found below ~50K then a linear decrease presents up to 390K, where both ZFC and FC branches overlap over the whole temperature range. The F-center exchange mechanism (FCE)[18,19] contributes to the ferromagnetism of KNLN-Fe. Electrons trapped inside the oxygen vacancy constitutes an F-center (two electrons in an oxygen vacancy). The exchange interaction between two $Fe^{3+}$ ions nearby via the F-center brings about the ferromagnetism.

In contrast to KNLN-Fe, KNLN-Cu ceramic shows diamagnetism with similar hysteresis loops at different temperatures, shown in Fig. 4(b1). The KNLN ceramic is also diamagnetic and can be seen in the inset part of Fig. 4(b1). It is clear that diamagnetic property was enhanced with the introduction of Cu ion. Therefore, a different exchange interaction between point defects from KNLN-Fe could contribute to the diamagnetic property. The temperature dependence of magnetization for KNLN-Cu ceramic at ZFC/FC in a temperature range from 15 to 390K is shown in Fig. 4(b2). The absolute value of magnetization increases rapidly with the increase of temperature below 100K then decreases slightly up to 390K, meanwhile, ZFC and FC curves do not overlap in the temperature range from 100 to 390K.

The X-band (9.4 GHz) EPR spectrum of KNLN and KNLN-Cu at 90K is shown in Fig. 5. A clear *g*-value and hyperfine splitting can be found in KNLN-Cu. The $Cu^{2+}$-center in KNLN-Cu X-band EPR is enough to resolve different spectral features. The two different $Cu^{2+}$-centers are approved by two stick-spectra. The high-field section of the spectrum is governed by overlapping transitions of different orientations originating from strong copper hyperfine splitting, and is further complicated by extra absorption peaks. The ordering of the observed g-values contributes to octahedrally coordinated $Cu^{2+}$ centers along the *z*-axis, which associates with a five-fold orbital degeneracy of the $3d^9$ ion that is split in the presence of an octahedral crystal field into a triplet ($t_{2g}$) and a doublet ($e_g$), with the latter lying lowest. A tetragonal distortion, caused by the crystal field and the Jahn-Teller effect, splits the $e_g$ levels further, resulting in an orbital-singlet $d_{x^2-y^2}$ ground state.

The results indicate that Cu preferentially substitutes at B site, furthermore, it can trap one or two oxygen vacancies. Two types of defect associates are formed $(Cu_{Nb}''' - V_O^{\bullet\bullet})'$ and $(V_O^{\bullet\bullet} - Cu_{Nb}''' - V_O^{\bullet\bullet})^{\bullet}$ [13], which is different from that of KNLN-Fe [18]. Since the two kinds of defect carry opposite charges, it is theoretically possible to keep overall charge neutrality by mutual compensation of the defect associates without formation of additional non-associated oxygen vacancies. For the latter type, the $Cu^{2+}$ ion should relax back into the plane spanned by the four equatorial oxygens [20] (which agrees with Jahn-Teller distortion). As a consequence, the distance to the two oxygen vacancies are approximately equal, meanwhile, the electric dipole moment for that defect association almost vanishes. The special defect structure is also the origin of double-hysteresis loop of KNLN-Cu.

The difference of magnetic performance between KNLN-Fe and KNLN-Cu should result from defect structure and ion characteristic. For the KNLN-Cu, the enhanced diamagnetism may be related to $Cu^+$ at A site by the diamagnetism of its $d^{10}$ shell instead to $Cu^{2+}$ is paramagnetic ions. Erünal [21] reported the coexistence of $Cu^+/Cu^{2+}$ presented in Cu-doped $K_{0.5}Na_{0.5}NbO_3$ ceramic based on EPR analysis. The sintering temperature didn't promote any new defect centers but higher sintering temperature promoted $Cu^{2+}$ oxidation state instead of $Cu^+$. In terms of defect chemistry, if the incorporation of $Cu^{2+}$ at the B-site is considered, $3/2$ $V_O^{\bullet\bullet}$ per Cu-ion will be introduced to the KNN lattice according to the incorporated reaction.

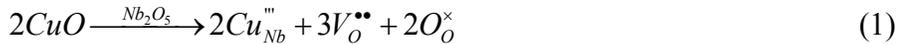

$$2CuO \xrightarrow{Nb_2O_5} 2Cu_{Nb}''' + 3V_O^{\bullet\bullet} + 2O_O^{\times} \qquad (1)$$

On the other hand, by assuming that $Cu^+$ incorporates isovalently at the A-site and annihilates $1/2\,V_O^{\bullet\bullet}$ per Cu-ion results in following reaction.

$$2CuO+V_O^{\bullet\bullet}\xrightarrow{Na_2O}2Cu_{Na}^{\bullet}+O_O^{\times}\quad\text{or}\quad 2CuO+V_O^{\bullet\bullet}\xrightarrow{K_2O}2Cu_K^{\bullet}+O_O^{\times}\qquad(2)$$

Subsequently, the $Cu^{2+}$ changes its oxidation state to $Cu^+$ by trapping electronic defect states as follows.

$$Cu_{Na}^{\bullet}+e'\rightarrow Cu_{Na}^{\times}\quad\text{or}\quad Cu_K^{\bullet}+e'\rightarrow Cu_K^{\times}\qquad(3)$$

The e′ in turn are generated during high-temperature processing by the following defect equilibrium

$$O_O^{\times}\Leftrightarrow V_O^{\bullet\bullet}+2e'+\frac{1}{2}O_2\uparrow\qquad(4)$$

Adopting a picture, after which lattice vacancies promote interatomic diffusion during calcination and sintering [22,23], the incorporation of $Cu^{2+}$ at the B-site increases the amount of $V_O^{\bullet\bullet}$ in the lattice and thus enables the formation of highest dense ceramics for KNLN-Cu (Table 1).

Cu$^{2+}$ ions with nine $d$-electrons in the valence shell ($t_{2g}$ and $e_g$) in an octahedral environment with six equivalent atoms in the first coordination sphere constitute a classic example of Jahn-Teller systems. For such Cu$^{2+}$ complexes, dynamic Jahn-Teller behavior is commonly observed at room temperature [24]. On lowering the temperature, the dynamic behavior is likely to vanish below a temperature called the Jahn-Teller transition temperature and a static distortion occurs. The distortions depend strongly on the matrix of the ligand as well as on the geometrical environment [25]. This can explain the $M$(T) curves of KNLN-Cu at ZFC/FC which overlap at low temperature but separate at high temperature.

Considering the coexistence of ferroelectric, diamagnetic and ferromagnetic orders in the KNLN-Fe and KNLN-Cu ceramics, one can expect the coupling between ferroelectricity and ferromagnetism in the system. Fig. 6 shows capacitance (C) and admittance (G) as a function of frequency at zero filed and H=2 kOe field for KNLN-Fe (Fig. 6a) and KNLN-Cu (Fig. 6b) ceramics. A slight increase of capacitance occurs in KNLN-Fe sample (Fig. 6a1) at low frequency while there is no change in KNLN-Cu sample (Fig. 6b1) in the whole frequency under magnetic field. Furthermore, KNLN-Fe seems to show a very weak magnetoresistance effect (Fig. 6a2) in the low frequency range. The effect remains when the magnetic filed is removed, indicating that it is irreversible. The effect of magnetic field on capacitance and resistance indicates the strong coupling between the ferroelectric and ferromagnetic orders in KNLN-Fe. But for KNLN-Cu, the coupling is not clear (Fig. 6b2) due to the special defect structure and Jahn-Teller effect of Cu ion. Therefore, the defect type and ion characteristic are the key role to control the coupling between the ferroelectric and ferromagnetic in perovskite ferroelectrics.

## 4. Conclusions

In summary, we experimentally proved the unusual ferroelectric and magnetic properties in the Cu-doped KNLN and Fe-doped KNLN ceramics. The magnetic hysteresis loops are clearly observed in the temperature range of 15 K to 390 K and present great difference between Fe-doped KNLN and Cu-doped KNLN. Further studies on ferroelectric, ferromagnetism and diamagnetism properties in other perovskite oxides are expected. The present experimental results are expected to open up new possibilities in the investigations of the physical mechanism and the multiferroic application for the perovskite oxides as well as spin electronics.

**Acknowledgements**

This work was supported by the Natural Science Foundation of China (No. 51002036). We thank Dr. Jun Lu and Prof. Yicheng Cai (Institute of Physics, Chinese Academy of Sciences) for high temperature magnetization measurements, and Dr. Badari Narayana Rao for the English writing.

**Figure Captions**

Table 1   Density, grain size and cell parameter of KNLN-base ceramics

Fig. 1   XRD patterns of (a) KNLN, (b) KNLN-Fe and (c) KNLN-Cu ceramics and results of crystallographic refinement using the space group *Pm*. Inset, SEM micrographs of polished and thermal etched surfaces of sintered KNLN, KNLN-Fe, KNLN-Cu, respectively. The quality of fit of the refinements is shown at the right bottom of the picture.

Fig. 2   Energy disperse spectroscopy (EDS) of (a) KNLN, (b) KNLN-Cu and (c) KNLN-Fe ceramics.

Fig. 2   Polarization hysteresis loops of the KNLN, KNLN-Fe and KNLN-Cu ceramics at room temperature and 1 Hz.

Fig. 3   (a) Field dependence of magnetization for the KNLN-Fe (a1) and KNLN-Cu (b1) ceramics at different temperatures. KNLN-Fe shows ferromagnetism while KNLN-Cu diamagnetism. Inset, field dependence of magnetization of the KNLN, shows weak diamagnetism. (b) Temperature dependence of magnetization for the KNLN-Fe (a2) and KNLN-Cu (b2) ceramics at zero-field cooled (ZFC) and H=0.5T field cooled (FC).

Fig. 4   The comparison of X-band measurements for pure KNLN and KNLN-Cu samples at 90 K. Center I which has a larger g-value and hyperfine splitting than center II is observed in the sample of KNLN-Cu.

Fig. 5   Capacitance (C) and admittance (G) as a function of frequency at zero filed and H=2 kOe field for KNLN-Fe (4a) and KNLN-Cu (4b) ceramics. A memorial magnetoresistance effect can be found in KNLN-Fe sample.

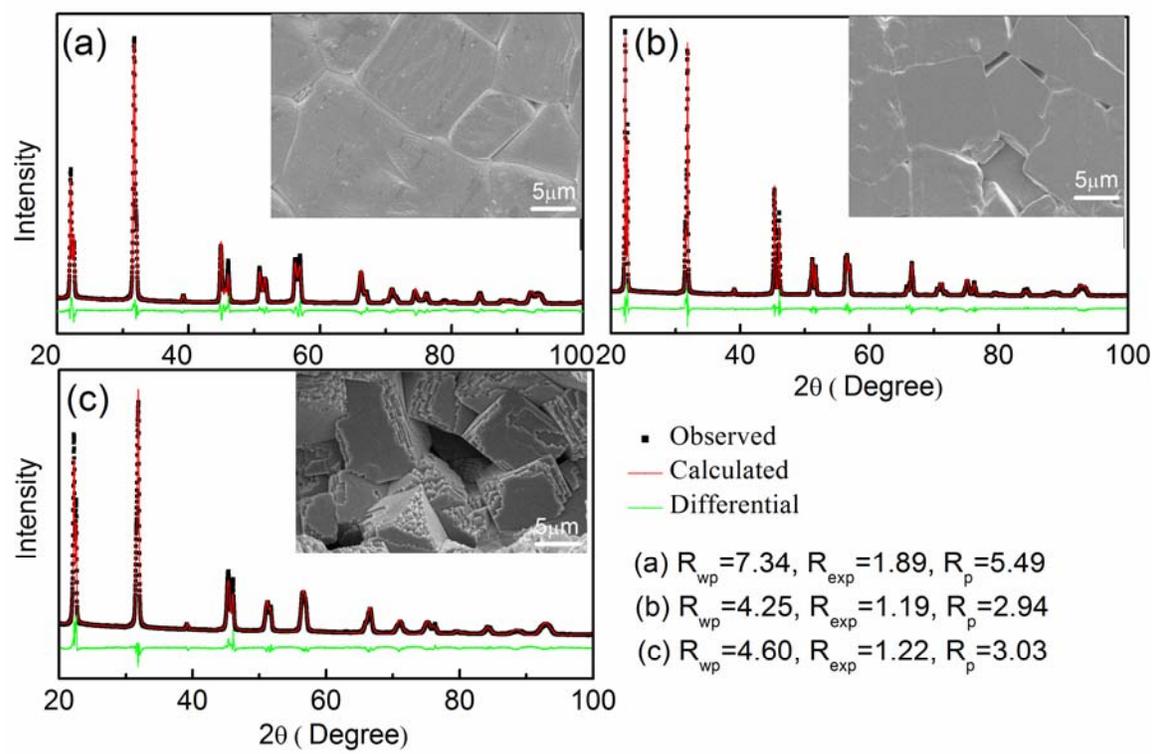

■ Observed
— Calculated
— Differential

(a) $R_{wp}$=7.34, $R_{exp}$=1.89, $R_p$=5.49
(b) $R_{wp}$=4.25, $R_{exp}$=1.19, $R_p$=2.94
(c) $R_{wp}$=4.60, $R_{exp}$=1.22, $R_p$=3.03

Fig. 1

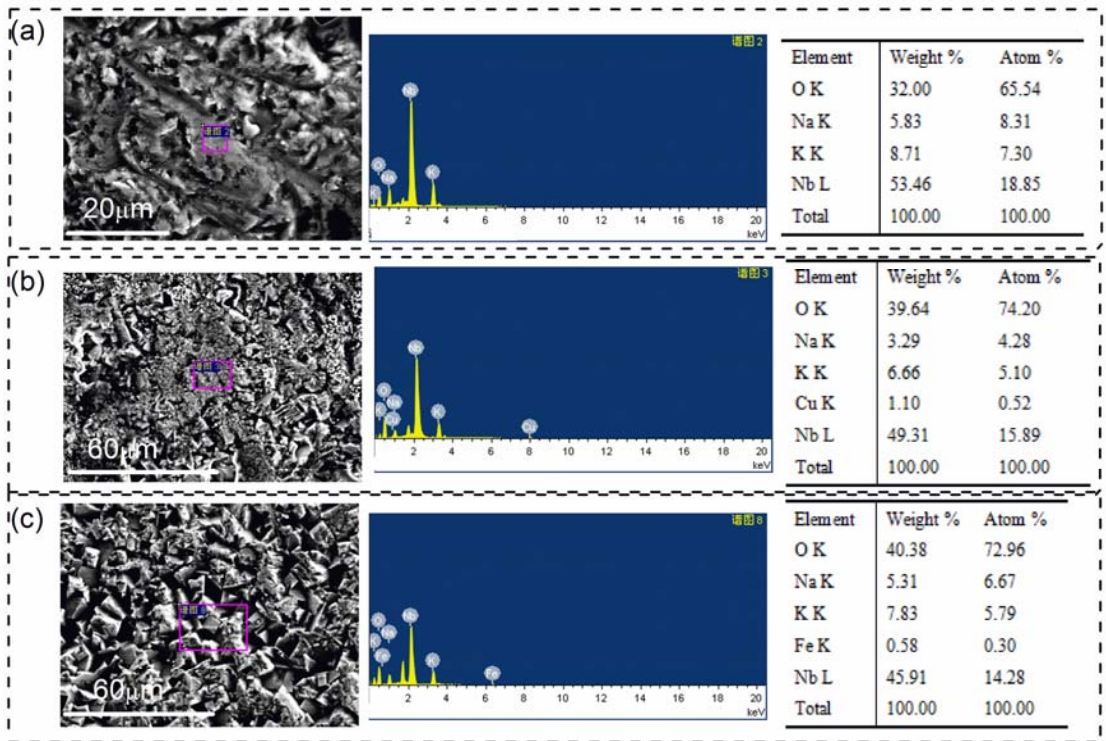

Fig. 2

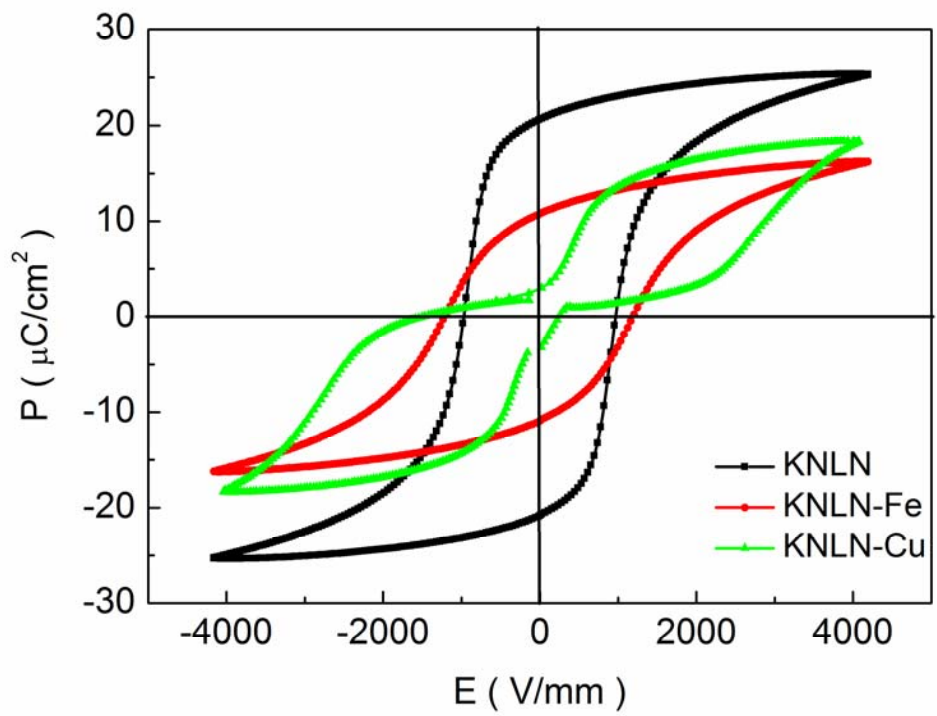

Fig. 3

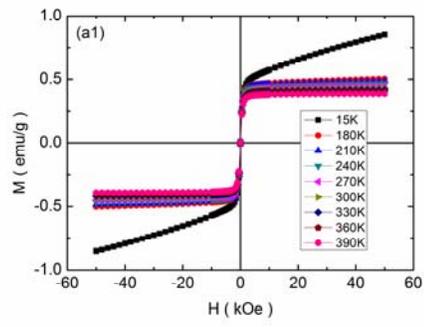
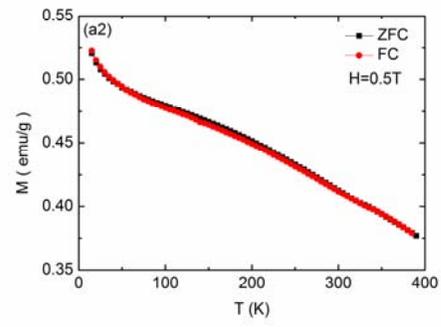

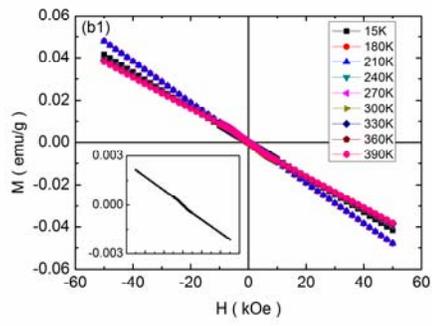
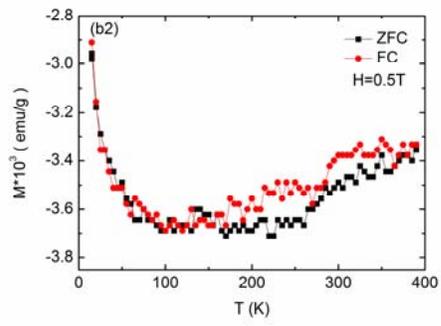

Fig. 4

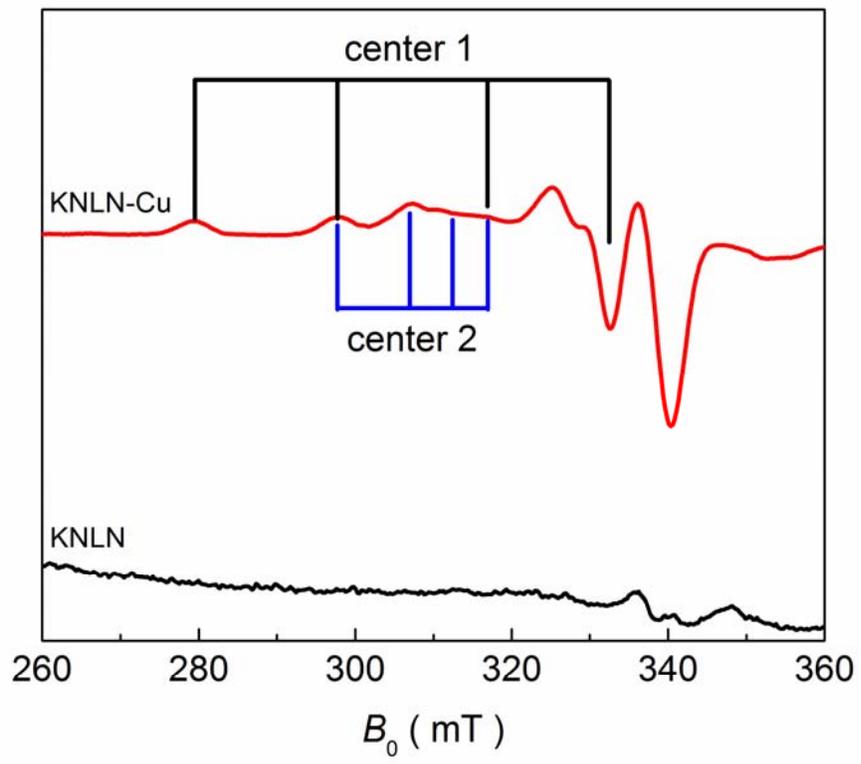

Fig. 5

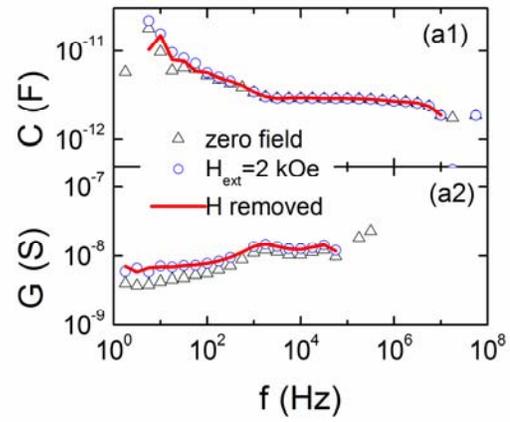

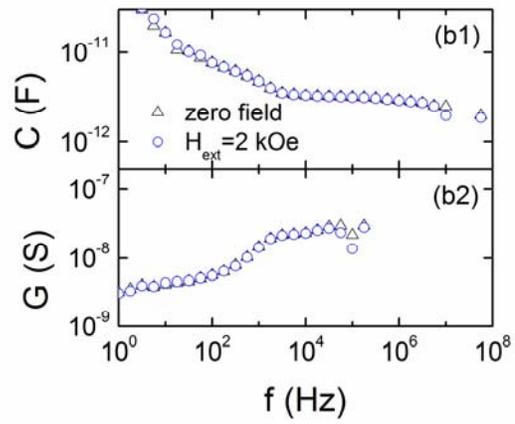

Fig. 6